\documentclass[aps]{revtex4}

\usepackage{amsmath}
\usepackage{graphicx}

\draft
\newcommand{\be}{\begin{equation}}
\newcommand{\ee}{\end{equation}}
\newcommand{\ba}{\begin{eqnarray}}
\newcommand{\ea}{\end{eqnarray}}

\begin{document}

\preprint{HEP/123-qed}
\title[Short title for running header]{On alternative approaches to Lorentz violation invariance
in loop quantum gravity inspired models}
\author{Jorge Alfaro$^{1}$$\footnote{jalfaro@puc.cl}$, Marat Reyes $^{1}$$\footnote{cmreyes@puc.cl}$, Hugo A. Morales-T\'ecotl$^2$$\footnote{Associate member of Abdus Salam International Centre for Theoretical
Physics, Trieste, Italy; hugo@xanum.uam.mx}$ and L.F. Urrutia$^3$}
\affiliation{1. Facultad de F\'{\i}sica,\\
Pontificia Universidad Cat\'olica de Chile\\ Casilla 306, Santiago 22, Chile\\
2. Departamento de F\'{\i}sica,\\ Universidad Aut\'onoma Metropolitana Iztapalapa \\
A.P. 55-534, M\'exico D.F. 09340, M\'exico \\
3. Departamento de F\'{\i}sica de Altas Energ\'{\i}as,\\ \normalsize
Instituto de Ciencias Nucleares, \\ \normalsize
Universidad Nacional Aut\'onoma de M\'exico,\\ \normalsize
A.P. 70-543, M\'exico D.F. 04510, M\'exico } 

\begin{abstract}
Recent claims point out that possible violations of  Lorentz
symmetry appearing   in some semiclassical models of extended
matter dynamics motivated by loop quantum gravity can be removed
by a different choice of canonically conjugated variables. In this
note we  show  that such alternative is inconsistent with the
choice of variables in the underlying quantum theory together with
the semiclassical approximation,  as long as the correspondence
principle is maintained. A consistent choice will violate standard
Lorentz invariance. Thus, to preserve a relativity principle in
this framework, the linear realizations of Lorentz symmetry should
be extended or superseded.

\vskip.2cm


\end{abstract}

\maketitle


Background independent nonperturbative quantum gravity in the form
of Loop Quantum Gravity has made some progress. Black hole entropy
\cite{BH}, non singular cosmological models \cite{BHAMT} and
Planck corrected effective matter dynamics \cite{GP,AMU,THIEMANN}
are amongst the problems that { recently} have received { a great
deal of} attention in this context. However, a key open problem is
that of defining good semiclassical states yielding a correct
semiclassical limit and yet allowing to calculate possibly new
quantum gravity effects. Now, heuristic semiclassical states
\cite{GP,AMU} and coherent states \cite{THIEMANN} both indicate
that modifications to the standard particle dynamics can appear in
the effective theory for matter and possibly gravity, in the form
of Planckian corrections which therefore were considered
unaccessible to either experiments or observation. The situation
has changed by considering high precision atomic experiments as
well as detection of gamma ray bursts together with high energy
cosmic rays, among other possibilities. The latter setting has
been proposed to test the modified dispersion relations for
particle propagation \cite{AC1}, that constitute the most direct
implication of such modified dynamics. Up to now, such effects
have been mostly interpreted as signaling the breakdown of
standard particle (active) Lorentz transformations
\cite{kostelecky1}, through the appearance of a privileged
reference frame, usually identified with that in which the cosmic
radiation background looks isotropic. Although the experimental
and observational bounds to such violations are very stringent
\cite{bounds}, there is the possibility of having an extension of
the relativity principle that comes to terms with the existence of
Planck scale modified particle dynamics. For instance, double
special relativity \cite{DSR} allows for the coexistence  of
deformed dispersion relations and extended transformations laws
relating inertial reference frames. This includes  non linear or
deformed  realizations of the Poincare algebra. This is a
possibility which deserves further investigation to elucidate its
physical realization in terms of particle interactions \cite{AC2}.

The  construction of the semiclassical approximation in Refs.
\cite{AMU,THIEMANN}  starts from a well defined quantum
Hamiltonian \cite{THIEMANN1}, where pairs of canonical variables
are established {\it ab initio} by providing the corresponding
commutation relations. Subsequently, heuristic semiclassical
states are proposed to define the effective Hamiltonian and the
classical limit is further achieved by using the correspondence
principle to identify the associated canonically conjugated
variables of the classical phase space.

In the case of the Gambini-Pullin (GP) electrodynamics \cite{GP},
there are recent claims in the literature \cite{CORDOBA} that it
is permissible to make a particular choice of classical canonical
variables in the  Planck scale modified GP Hamiltonian, which
leads to the standard Lorentz covariant Maxwell equations instead
of the  modified ones \cite{GP,AMU}. In this note we show that the
quantum origin of the canonical variables, together with the
standard correspondence principle, does not allow to consistently
consider further redefinitions at the classical level, with the
exception of canonical transformations.

To make our point clear let us consider, instead of the
electromagnetic Hamiltonian written in loop quantum  variables
\cite{THIEMANN1}, the simplest case of a quantum harmonic
oscillator in the usual representation.
\begin{eqnarray}
{\cal H}_{\rm Quant} &=&  \frac{1}{2} (\hat{P}^2 + \hat{Q}^2), \label{eq:qh}\\
 \left[\hat{Q} , \hat{P}\right]  &=& \imath\, \hbar.
\end{eqnarray}
In analogy to
what is done in Refs.\cite{AMU,THIEMANN} we choose a semiclassical
state, here given by the coherent state $|z\rangle$ \cite{MERZBACH},
\begin{equation}
{\hat a}\,|z\rangle=z\,|z\rangle,\quad \langle z|\,{\hat a}^{\dagger}=\langle z|\,z^*,\quad
z=\frac{1}{\sqrt{2}}(\alpha+\imath\,\beta), \quad  {\hat a}=\frac{1}{\sqrt{2}}(\hat{Q}+i\hat{P})
\end{equation}
and define the classical effective Hamiltonian $H_{class}$ as the
expectation value of (\ref{eq:qh}) in $|z\rangle$, so that
\begin{equation}
H_{class}=\langle z|\,{\cal H}_{\rm Quant} \,|z\rangle
\end{equation}
Then
\begin{eqnarray}
\langle z|\hat{Q}| z \rangle &=& \alpha, \qquad \langle z|\hat{P}|
z \rangle = \beta, \qquad  \langle z|{\hat a}^\dagger\, {\hat a}| z \rangle=
|z|^2,\nonumber \\
H_{class}(\alpha,\beta)&=& |z|^2=\frac{1}{2}
\left[\alpha^2+\beta^2\right]\label{eq:ch}
\end{eqnarray}
where we { have} neglected the zero point energy term. An
arbitrary choice of canonical variables $q=q(\alpha,\beta),
p=p(\alpha,\beta)$ dictated by some additional criteria could lead
to almost any dynamics for the variables $\alpha, \, \beta$, and
would be inconsistent with the idea of a classical limit
descending from a specified quantum theory { via the standard
correspondence principle}. For example, the choice of canonical
variables
\begin{equation}
p=\alpha^2+\beta^2, \qquad q=\beta^2,\qquad \{q,p\}=1
\end{equation}
leading to
\begin{equation}
\{\alpha,\beta\}\neq1,
\end{equation}
defines a dynamics for the variables $\alpha, \beta$ which is clearly not
related by a canonical transformation to
the oscillator problem.

Certainly we expect the correspondence
\begin{equation}
\langle z|\hat{Q}| z \rangle \rightarrow q , \qquad  \langle z|\hat{P}| z \rangle
\rightarrow p,\quad  \frac{1}{\imath\hbar} \left[\hat{Q} , \hat{P}\right] \rightarrow
\left\{q,\,p \right\},
\end{equation}
to hold in order to have a sensible classical limit.
Hence one cannot but identify
\begin{eqnarray}
q:=\alpha, \quad p:=\beta, \quad \left\{ \alpha,\beta\right\} = 1,
\label{eq:aqbp}
\end{eqnarray}
to ensure the quantum Hamiltonian, Eq. (\ref{eq:qh}), has as its semiclassical
limit (\ref{eq:ch}). An arbitrary choice
\begin{eqnarray}
q=f(\alpha, \beta), \qquad p=g(\alpha, \beta), \qquad  \left\{p,
q\right\} &=& 1,
\end{eqnarray}
will not yield the right semiclassical behavior unless the
functions $f,g\,$ provide a canonical transformation from the
choice (\ref{eq:aqbp}).

In the loop quantum gravity expression for the electromagnetic
quantum Hamiltonian, the conjugated pair of electromagnetic
variables defining the theory is $ \hat{A}_i({\vec x}),
\hat{E}^j(\vec y)$ , satisfying \cite{GP,AMU,THIEMANN,THIEMANN1}
\begin{equation}
\left[ \hat{A}_i({\vec x}), \hat{E}^j(\vec y)\right]=i \hbar \delta_i^j\,
\delta(\vec{x}-\vec{y}).
\label{COMM}
\end{equation}
Also we have $\hat{B}_i=\epsilon_{ijk}\partial_j\,\hat{A}_k$,
which is equivalent to ${\hat F}=d{\hat A} \,$ defined  on the 3-D
spatial surface of the canonical formalism. In the Hamiltonian
formulation, once we start with the basic commutator (\ref{COMM}),
it is  inconsistent to demand
\begin{equation} {\hat E}_i= \partial{\hat A}_i/ \partial t,
\label{DADT}
\end{equation}
as proposed in \cite{CORDOBA}. In fact $\partial{\hat A}_i/ \partial t$
arises from the  equation of motion
\begin{equation}
i\hbar\,\partial{\hat A}_i/ \partial t=[{\hat A}_i, {\hat H } ],
\qquad {\hat H}=\int (d^3x) {\hat {\cal H}}
\end{equation}
Moreover, the  expression for $ {\hat F}_{0i}$, which completes
the four-dimensional expression  ${\hat F}=d{\hat A},\, {\hat
A}={\hat A}_\mu dx^\mu,\, \mu=0,1,2,3$ comes as a consequence of
the equations of motions for ${\hat E}_i$, together with the fact
that ${\hat A}^0$ is the Lagrange multiplier associated with the
constraint $\partial_i{\hat E}^i=0$ in the vacuum situation.

Under the semiclassical
state $|W,\vec{E}, \vec{B}\rangle$ we have
\begin{eqnarray}
&&\langle W,\vec{E}, \vec{B}| \dots \hat{E}^i\dots |W,\vec{E}, \vec{B}\rangle
\rightarrow E^i,\nonumber \\
&&\langle W,\vec{E}, \vec{B}| \dots \hat{A}_i\dots |W,\vec{E},
\vec{B}\rangle \rightarrow A_i,
\end{eqnarray}
which leads to the modified vacuum Hamiltonian density
\begin{equation}
{\cal H}_{class}=\frac{1}{2}\left({\vec E}^2+{\vec B}^2\right)+ \alpha \, \ell_P
\left({\vec E}\cdot\nabla\times{\vec E} +
{\vec B}\cdot\nabla\times{\vec B}\right) + A^0\, \partial_i\,E^i,
\label{HAM}
\end{equation}
together with the non-zero Poisson brackets
\begin{equation}
\left\{{A}_i({\vec x}), {E}^j(\vec y)\right\}= \delta_i^j\,\delta(\vec{x}-\vec{y}),
\end{equation}
obtained from (\ref{COMM}) via the correspondence principle. The
above canonical structure leads to modified Maxwell's equations.
We recover the four dimensional version of $F=dA$, but the
equations of motion now yield
\begin{equation}
F_{0i}=\left(\vec E + \alpha\,\ell_P \nabla\times {\vec E} \right)_i \, ,
\label{FOI}\end{equation}
instead of the
standard situation, which is recovered in the limit $\ell_P
\rightarrow 0$. The relation (\ref{FOI}) can  also be  obtained directly from the modified
Maxwell's equations.

Summarizing, we have shown that the choice (\ref{DADT}) is
inconsistent with the starting point (\ref{COMM}) of the
Hamiltonian formulation of LQG. Also,  any other choice of the
classical canonical variables, up to canonical transformations,
looks highly artificial and it would be inconsistent with the
underlying quantum theory, provided the correspondence principle
is maintained.

Of course one can take the approach of Ref.\cite{CORDOBA} starting
from the Gambini-Pul\-lin Hamiltonian  (\ref{HAM}) and
constructing a canonical momentum $\pi^i=\pi^i(E)$ such that $
E_i=\partial A_i / \partial t=-\partial H/ \partial \pi^i $. The
problem here would be  to identify the regularized  quantum
gravity  theory corresponding to such choice. Some preliminary steps towards
the construction of a manifestly Lorentz invariant loop quantization of gravity
appear in Ref.\cite{alexandrov}.

Although we have clarified  the differences arising when
arbitrarily  reassigning canonical variables to a semiclassical
limit descending from a given quantum theory in the context of
Lorentz violating electrodynamics,  according to the proposals
\cite{CORDOBA}, the same line or argument can be generalized,
mutatis mutandis, to the case of spin-$\frac{1}{2}$ particles. On
top of the afore mentioned problems there is the well known result
that even classically equivalent theories are not necessarily
equivalent at the quantum level. This latter point should not be
confused with the argument of the present note.

In closing we stress that Planckian symmetry violations of the
kind derived so far in Ref. \cite{GP,AMU,THIEMANN} are a property
of the proposed semiclassical states and, moreover, rather than
indicating the existence of a privileged observer, they might be
elucidating an unknown symmetry superseding the linear realization
of the Lorentzian one.

\section*{Acknowledgments}

LFU thanks R. Montemayor for useful comments in relation to this
problem, together with support from the projects CONACYT-40745-F
and DGAPA-IN11700. CMR acknowledges support from project (Apoyo de
T\'esis Doctoral) CONICYT (Chile), and wants to thanks professors
L. F. Urrutia and H. A.  Morales-T\'ecotl for their hospitality at
ICN and UAMI. The work of JA has been partially supported by
Fondecyt 1010967. HMT acknowledges partial support from the
project CONACYT-40745-F.

\end{document}